\documentclass[a4paper,twoside]{article}

\usepackage{epsfig}
\usepackage{subcaption}
\usepackage{calc}
\usepackage{amssymb}
\usepackage{amstext}
\usepackage{amsmath}
\usepackage{amsthm}
\usepackage{multicol}
\usepackage{pslatex}
\usepackage{natbib} 
\let\bibhang\relax  
\usepackage{apalike}

\usepackage[T1]{fontenc}
\usepackage[noend]{algpseudocode}
\usepackage[dvipsnames,svgnames]{xcolor}
\usepackage[linesnumbered,ruled,vlined]{algorithm2e}
\usepackage{amsfonts}
\usepackage{array}
\usepackage{appendix}
\usepackage{bm}
\usepackage{booktabs}
\usepackage{caption}
\usepackage{dirtytalk}
\usepackage{graphicx}
\usepackage{hyperref}
\usepackage{multirow}
\usepackage{subcaption}
\usepackage{tabularx}
\usepackage{xfrac}
\usepackage{paralist}
\usepackage{wrapfig}
\usepackage{tikz}
\usepackage{geometry}
\usepackage{fleqn}
\usepackage{float}
\usepackage{afterpage}
\usepackage{longtable}
\usepackage{hyperref}
\hypersetup{
  colorlinks=true,           
  allcolors=blue!70!black,   
  pdfstartview=Fit,          
  breaklinks=true,
  pdfauthor={H. Alshareef, S. Stucki and G. Schneider},
  pdftitle={Transforming Data Flow Diagrams for Privacy Compliance (Long Version)}}
\SetKwInOut{Input}{input}
\SetKwInOut{Output}{output}
\theoremstyle{definition}
\newtheorem{definition}{Definition}

\newcommand{\ie}{i.e.,\ }
\newcommand{\eg}{e.g.,\ }

\makeatletter
\newcommand\footnoteref[1]{\protected@xdef\@thefnmark{\ref{#1}}\@footnotemark}
\makeatother

\newcommand{\Activators}[1][]{\ensuremath{\mathcal{N}_{#1}}\xspace}
\newcommand{\Flows}[1][]{\ensuremath{\mathcal{F}_{#1}}\xspace}
\newcommand{\PADFD}{\ensuremath{\text{PA-DFDs}}\xspace}
\newcommand{\DFD}{\ensuremath{\text{B-DFDs}}\xspace}
\newcommand{\Types}[1][]{\ensuremath{\mathcal{T}_{#1}}\xspace}

\newcommand{\Values}[1][]{\ensuremath{\mathcal{V}_{#1}}\xspace}
\newcommand{\Attributes}[1][]{\ensuremath{\mathcal{A}_{#1}}\xspace}

\newcommand{\type}[1][]{\ensuremath{\mathit{type}}\xspace}
\newcommand{\partner}[1][]{\ensuremath{\mathit{partner}}\xspace}

\newcommand{\DataNodeTypes}{\Types[\mathsf{dn}]}
\newcommand{\PolNodeTypes}{\Types[\mathsf{pn}]}
\newcommand{\AdminNodeTypes}{\Types[\mathsf{an}]}
\newcommand{\RawFlowTypes}{\Types[\mathsf{rf}]}
\newcommand{\DataFlowTypes}{\Types[\mathsf{df}]}
\newcommand{\DataFlowTypesPA}{\Types[\mathsf{df}]^{P}}
\newcommand{\PolFlowTypes}{\Types[\mathsf{pf}]}
\newcommand{\AdminFlowTypes}{\Types[\mathsf{af}]}

\newcommand{\nAtt}[1][\Activators]{\ensuremath{\ell_{#1}}\xspace}
\newcommand{\eAtt}[1][\Flows]{\ensuremath{\ell_{#1}}\xspace}

\newcommand{\externalentity}{\ensuremath{\mathit{ext}}\xspace}
\newcommand{\process}{\ensuremath{\mathit{proc}}\xspace}
\newcommand{\database}{\ensuremath{\mathit{db}}\xspace}
\newcommand{\limit}{\ensuremath{\mathit{limit}}\xspace}
\newcommand{\request}{\ensuremath{\mathit{request}}\xspace}
\newcommand{\reason}{\ensuremath{\mathit{reason}}\xspace}
\newcommand{\logg}{\ensuremath{\mathit{log}}\xspace}
\newcommand{\logdb}{\ensuremath{\mathit{log\_db}}\xspace}
\newcommand{\clean}{\ensuremath{\mathit{clean}}\xspace}
\newcommand{\policydb}{\ensuremath{\mathit{policy\_db}}\xspace}
\newcommand{\pf}{\ensuremath{\mathit{pf}}\xspace}
\newcommand{\df}{\ensuremath{\mathit{df}}\xspace}
\newcommand{\into}{\ensuremath{\mathit{in}}\xspace}
\newcommand{\out}{\ensuremath{\mathit{out}}\xspace}
\newcommand{\comp}{\ensuremath{\mathit{comp}}\xspace}
\newcommand{\store}{\ensuremath{\mathit{store}}\xspace}
\newcommand{\readd}{\ensuremath{\mathit{read}}\xspace}
\newcommand{\delete}{\ensuremath{\mathit{delete}}\xspace}

\newcommand{\inaddnewelement}{\ensuremath{\mathit{addInElems}}\xspace}
\newcommand{\outaddnewelement}{\ensuremath{\mathit{addOutElems}}\xspace}
\newcommand{\compaddnewelement}{\ensuremath{\mathit{addCompElems}}\xspace}
\newcommand{\storeaddnewelement}{\ensuremath{\mathit{addStoreElems}}\xspace}
\newcommand{\readaddnewelement}{\ensuremath{\mathit{addReadElems}}\xspace}
\newcommand{\deleteaddnewelement}{\ensuremath{\mathit{addDeleteElems}}\xspace}
\newcommand{\addnewcomelement}{\ensuremath{\mathit{addCommonElems}}\@\xspace}
\newcommand{\sourcesactivators}{\ensuremath{S}\xspace}
\newcommand{\targetsactivators}{\ensuremath{T}\xspace}

\newcommand{\setsources}{\sourcesactivators(G)}
\newcommand{\settargets}{\targetsactivators(G)}

\DeclareMathOperator{\pto}{\rightharpoonup}

\newcommand{\nodeT}[2]{#1 \colon #2}
\newcommand{\edgeT}[4][]{#2 \colon #3 \rightsquigarrow_{#1} #4}
\newcommand{\xdasharrow}[2][->]{
\tikz[baseline=-\the\dimexpr\fontdimen22\textfont2\relax]{
\node[anchor=south,font=\scriptsize, inner ysep=1.5pt,outer xsep=2.2pt](x){#2};
\draw[shorten <=3.4pt,shorten >=3.4pt,dashed,#1](x.south west)--(x.south east);
}
}

\newcommand{\isSrc}[2][G]{#2 \in \sourcesactivators(#1)}
\newcommand{\isTgt}[2][G]{#2 \in \targetsactivators(#1)}

\newcommand{\gitrepo}{\url{https://github.com/alshareef-hanaa/PA-DFD-Paper}}
 
\usepackage{SCITEPRESS}

\begin{document}
\newenvironment{example}{\bgroup\par\noindent\textbf{Example}\par
\[}{\]\egroup}
\pagestyle{plain}

\title{Transforming Data Flow Diagrams for Privacy Compliance\\
  (Long Version\thanks{
    This is an extended version of a paper to be presented at
    MODELSWARD~2021.
    It contains a more detailed description of our transformation
    algorithm and an additional case study, which were not included in the
    conference paper.
  })}

\author{
  \authorname{Hanaa Alshareef\sup{1}, 
    Sandro Stucki\sup{2}\orcidAuthor{0000-0001-5608-8273} and
    Gerardo Schneider\sup{2}\orcidAuthor{0000-0003-0629-6853}}
\affiliation{\sup{1}Chalmers University of Technology, Gothenburg, Sweden}
\affiliation{\sup{2}University of Gothenburg, Gothenburg, Sweden}
\email{hanaa@chalmers.se, sandro.stucki@gu.se, gersch@chalmers.se} }


\abstract{Recent regulations, such as the European General Data Protection Regulation (GDPR), put stringent constraints on the handling of personal data.
Privacy, like security, is a non-functional property, yet most software design tools are focused on functional aspects, using for instance Data Flow Diagrams (DFDs).
In previous work, a conceptual model was introduced where DFDs could be extended into so-called Privacy-Aware Data Flow Diagrams (PA-DFDs) with the aim of adding specific privacy checks to existing DFDs.
In this paper, we provide an explicit algorithm and a proof-of-concept implementation to transform DFDs into PA-DFDs.
Our tool assists software engineers in the critical but error-prone task of systematically inserting privacy checks during design (they are automatically added by our tool) while still allowing them to inspect and edit the PA-DFD if necessary.
We have also identified and addressed ambiguities and inaccuracies in
the high-level transformation proposed in previous work. We apply our approach to two realistic applications from the construction and
online retail sectors.}

\onecolumn \maketitle \normalsize \setcounter{footnote}{0} \vfill

\section{\uppercase{Introduction}} \label{sec:introduction}

\noindent
The \emph{European General Data Protection Regulation} (GDPR) has been in place for more than two years now.
It
imposes stringent constraints on how individuals' personal data is to be collected and processed, stipulating heavy penalties in case of violations \citep{gdpr2016}.
As a consequence, public and private companies have been updating their privacy policies informing users how their data is being used.
Yet, it remains unclear whether the GDPR has had a substantial impact on the practices used by companies when handling personal data, not least because of the technical difficulty in complying with many of the regulation's clauses.
Implementing the right to be forgotten, for instance,
affects legacy storage media where data has been collected and third parties that have previously published personal data
\citep{politou2018forgetting,rubinstein2012big}.
Software engineers trying to meet the required data protection
principles often face a conflict between system and privacy requirements \citep{oetzel2014systematic}.

Barring such trade-offs, privacy-compliance is an ambitious goal in and by itself.
When talking about privacy one does not refer to one particular property but rather to a set of properties, including well-known security properties like confidentiality and secrecy, as well as other concepts like data minimisation (DM), privacy impact assessment (PIA), user consent, the right to be forgotten, purpose limitation, etc.
But even when restricted to a specific privacy property, verifying the privacy compliance of legacy software remains a very difficult task---the problem is in general undecidable \citep{TBC15pbd,schneider18pbc}.
We therefore advocate an alternative approach
known as the {\em Privacy by Design} (PbD) principle \citep{cavoukian2011privacy}.
PbD says, roughly, that any (computerised) personal data processing environment should be designed taking privacy into account from the very beginning of the (software) development process---even as early as the requirement elicitation phase.
It has been argued that PbD is more tractable than retrofitting 
legacy software for privacy compliance \citep[see e.g.][]{enisa15pdp}.

Still, the implementation of privacy principles such as PbD, PIA or DM requires a lot of work from software engineers and developers.
Several recent studies suggest that software engineers consider such principles to be overly complicated and impractical
and that they lack the necessary knowledge to implement them \citep{senarath2018developers,sirur2018we,freitas2018gdpr}.
As a consequence, many applications are designed with limited or no privacy considerations \citep{aktypi2017unwinding,ayalon2017developers}.
Hence, despite having been advocated since the mid-1990s, PbD has gained momentum only in recent years,
mostly due to the GDPR.
Progress has been made, in particular, in the development of methods for PbD (cf.\ Section~\ref{sec:relatedw}).

An example is the work by~\citet{ASS16pac,ASS18pcm},
who propose an approach to automatically add privacy checks already at the design level.
The updated design then 
guides software engineers in the implementation of privacy mechanisms.
The idea is based on model transformations, enhancing \emph{Data Flow Diagrams} (DFDs) with checks for specific privacy concepts, notably concerning retention time and purpose limitation for each operation on sensitive (personal) data (storage, forwarding, and processing of data).
The enhanced diagram is called a {\em Privacy-Aware Data Flow Diagram} (or PA-DFD for short).
The ultimate goal of that proposal is that the software engineer designs a DFD for the problem under consideration, pushes a button to obtain a PA-DFD, inspects it manually, and then generates a program template from the PA-DFD  that guides the programmer in the concrete implementation of the privacy checks.
\citeauthor{ASS16pac} describe their transformation from DFDs to PA-DFDs through high-level graphical ``rules'' but provide neither a full algorithm nor a reference implementation.  The main purpose of our paper is to provide these missing pieces.

Concretely, we make the following contributions.
\begin{compactenum}[i)]
\item We describe a pair of algorithms to check and automatically transform DFDs into PA-DFDs (previous work only gave a high-level graphical transformation). While defining the algorithms we identified some ambiguities and inaccuracies in the description given in the hotspots' translation by \citet{ASS16pac,ASS18pcm}. (Section \ref{sec:algorithms}).
\item We provide a Python implementation of our algorithms,\footnote{The sources are available at\\\gitrepo{}.} which processes DFD diagrams in an XML format compatible with the popular \texttt{draw.io} platform (Section \ref{sec:algorithms}).
\item We evaluate our algorithms on two case studies: an automated payment system
  and an online retail system (Section \ref{sec:casestudy}).
\end{compactenum}
We recall necessary background in the next section.

\section{\uppercase{Preliminaries}} \label{sec:Preliminaries}

\noindent We recall here relevant GDPR concepts, the definition of DFDs, as well as the transformation into PA-DFDs given by \citet{ASS18pcm}.

\subsection{GDPR} \label{subsec:GDPR}

The European {\em General Data Protection Regulation} (GDPR) 
contains 99~articles regulating \emph{personal data} processing.
The GDPR is organised around a number of key concepts, most notably its seven \emph{principles} of personal data processing, the {\em rights} of data subjects and six {\em lawful grounds} for data processing operations.
Relevant to this paper are the principles of \emph{purpose limitation} and \emph{accountability}, the lawful ground of \emph{consent}, and the \emph{right to be forgotten}.
The paper also indirectly touches on the right to information, access and rectification of personal information, and object to personal data processing.
See the regulation \citep{gdpr2016} and the critical review by \citet{HP16ngd} for more details on the GDPR.

\subsection{Data Flow Diagrams (DFDs)} \label{subsec:DFDs}

\begin{figure}
\centering
\includegraphics[width=0.99\columnwidth]{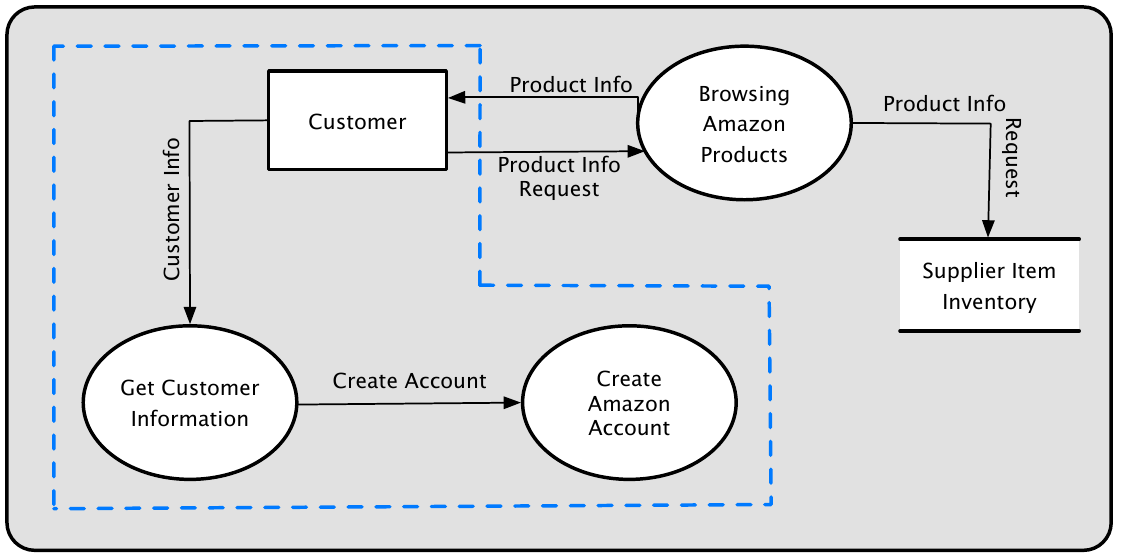}
\caption{Example of a DFD: High-level design of part of the e-store Ordering System.}
\label{fig:dfd}
\end{figure}

\noindent A {\em data flow diagram} (DFD) is a graphical representation of how data flows among different software components. As shown in Fig.~\ref{fig:dfd}, DFDs are composed of {\em activators} and {\em flows}. Activators can be \emph{external entities} (``boxes'' representing for instance end users and 3rd party systems), \emph{processes} (computation applied to the data in the system) and \emph{data stores}.
Processes may represent detailed low-level operations or be high-level, representing complex functionality that could be refined into sub-processes;
such \emph{complex} processes are represented by a double-lined circle or ellipse.
The flow of data is represented by \emph{data flow} arrows.
DFDs are subject to certain \emph{well-formedness rules} \citep{falkenberg1991}. For instance, activators cannot be isolated (disconnected from all other activators), direct data flow between two external entities or two data stores is not allowed, processes 
must have incoming and outgoing flows, etc.  We chose DFDs as the basis of our approach because they are a widely used tool for modelling digital systems. Furthermore, DFDs are commonly used for security and privacy analysis in software systems \citep{shostack2014threat,wuyts2014empirical}.

\citet{ASS16pac,ASS18pcm} extended the standard notation of DFDs with \emph{data deletion} type of flow, to indicate specific piece of data is to be deleted from a database.
They also added a data structure to specify information concerning personal data flow: (i) the owner of personal data, (ii) the purpose for the use of such data with an explicit consent from the data subject, and (iii) the retention time for the personal data (for how long the data may be used). This extension is referred to as \emph{Business-oriented DFD} (B-DFD) (see an example
in the top of Fig.~\ref{fig:DFDAmazon}).

\subsection{Adding Privacy Checks to DFDs} \label{subsec:privacy_DFDs}

\begin{figure}[tb]
\centering
\includegraphics[width=0.99\linewidth]{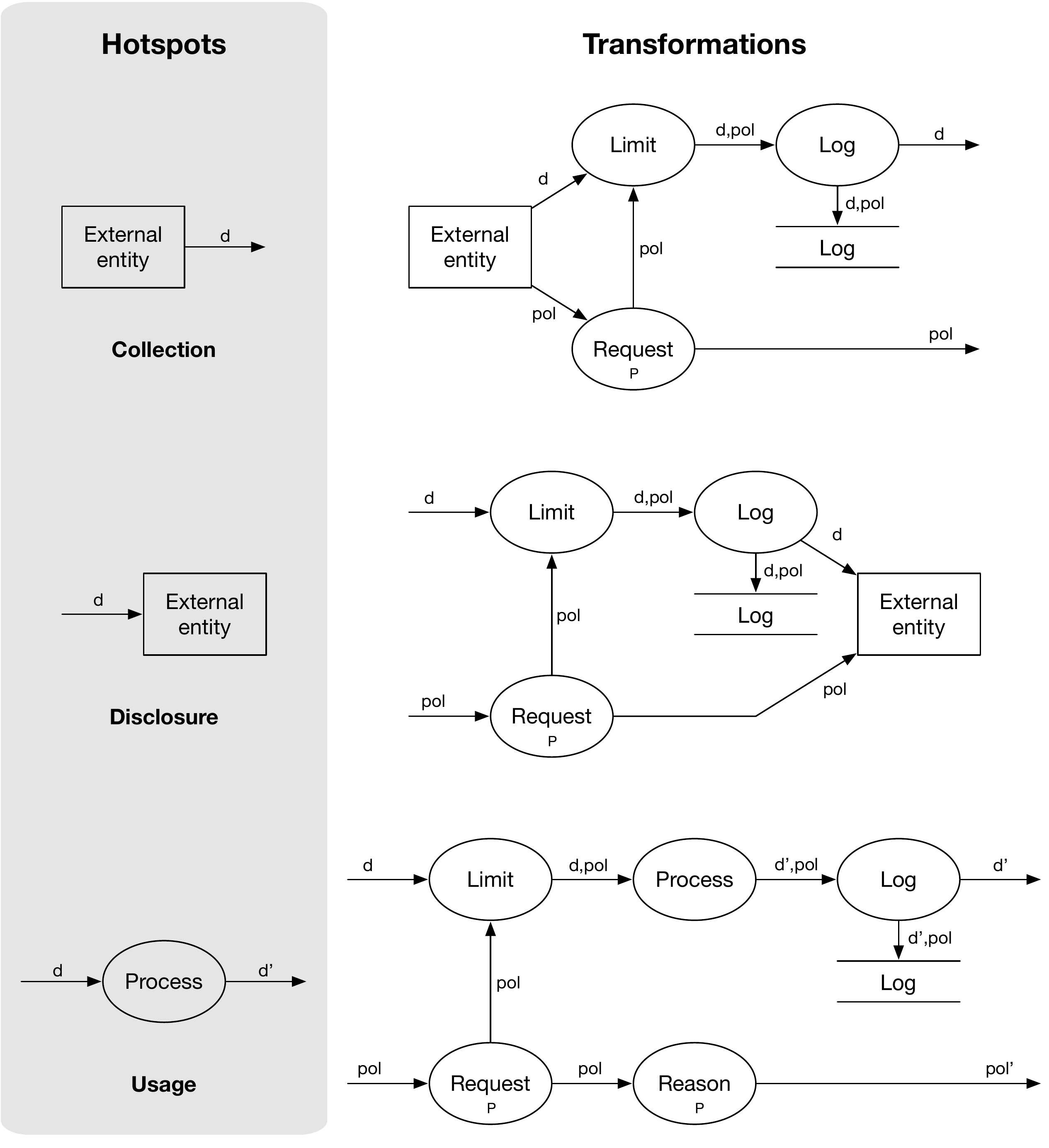}
\caption{Selection of B-DFD hotspots and corresponding PA-DFD elements \citep{ASS18pcm}.}
\label{fig:hotspots}
\end{figure}

The aim of the work by \citet{ASS16pac,ASS18pcm} is to (automatically) add privacy checks to an existing DFD (or rather to a B-DFD) to obtain a \emph{Privacy-Aware Data Flow Diagram} (PA-DFD) which contains relevant privacy checks for purpose limitation and retention time, as well as to ensure that everything is logged (for accountability) and to allow for users' policy management. \citeauthor{ASS18pcm} defined so-called \emph{hotspots} in the B-DFD to allow this transformation to be performed in a compositional way. The B-DFD hotspots and their corresponding PA-DFD elements are shown in Fig.~\ref{fig:hotspots}.

The left-hand side of Fig.~\ref{fig:hotspots} shows three types of hotspots. Each hotspot is defined by a pattern of activators and flows that corresponds to a basic data processing operation, such as ``collection'', ``disclosure'', etc.\ which is subject to the GDPR.  The notion of B-DFD hotspot thus provides a graphical means to identify personal data processing events in a system and to automatically introduce mechanisms to check and enforce the associated privacy policies.
The right-hand side of Fig.~\ref{fig:hotspots} shows, for each privacy hotspot in the B-DFD, the corresponding PA-DFD obtained by introducing a set of specialised activators and flows representing these privacy mechanisms.

Tables~1 and~2 in \cite{ASS18pcm} describes the privacy properties of interest for each hotspot, derived from the GDPR.
In order to capture the (new) privacy checks and to facilitate the transformation, the set of activator types in PA-DFDs is augmented with five subtypes of the ``Process'' activator type of B-DFDs: ``Limit'', ``Reason'', ``Request'', ``Log'' and ``Clean''.
Each of which corresponds to a particular privacy enforcement mechanism.
The ``Limit'' inspects whether the purpose of data processing is compatible with the data subject consent.  This limitation, in turn, demands a policy from the data subject, which is given by ``Request''. To provide information about data processing events in the context of their policies, ``Log'' is used to create log files in a Log data store. The ``Reason'' activator is used to get an updated policy corresponding to a newly computed data value. Finally, the ``Clean'' activator guarantees that personal data is eliminated from the data store.

The original description of PA-DFDs further allowed the new process elements to be tagged with a priority flag (`p') to indicate that their execution takes precedence over that of other activators.  This flag is not relevant to our work and we will ignore it in the following.
See \cite{ASS18pcm} for the rationale behind the above design decisions and for more explanations on the role of each specific element.

\begin{figure}[tb]
\centering
\includegraphics[width=\linewidth]{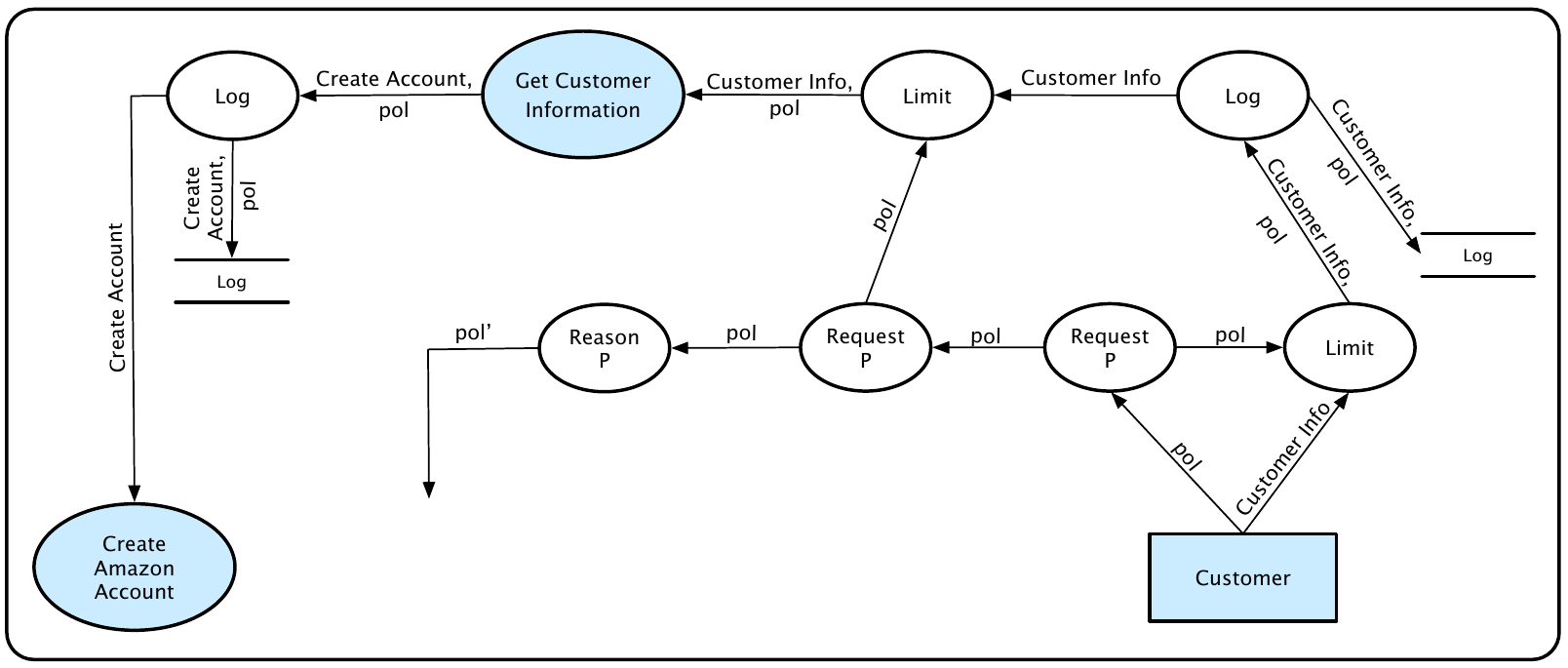}
\caption{Example of a PA-DFD generated by the old transformation rules}
\label{fig:example_3}
\end{figure}

To illustrate the transformation, consider Fig.~\ref{fig:example_3} where we show a B-DFD and its corresponding PA-DFD obtained by applying \citeauthor{ASS18pcm}'s rules.  In the figure, ``pol'' is a policy related to data ``d''.
Two rules (collection and usage) have been applied to a subset of the B-DFD from Fig.~\ref{fig:dfd}. The collection gets the personal data ``Customer Info'' and its corresponding policy ``pol'' from external entity ``Customer''. Then, the personal data flows to the
``Limit'' process that restricts data processing to the purposes that the data subject of ``d'' (in this example ``Customer Info'') has provided their consent to. The consent is specified in the policy ``pol'', received via the ``Request'' process.
``pol'' and ``d'' are logged by the ``Log'' process in the ``Log'' store where ``d'' is associated with its corresponding ``pol''.

Note, that the PA-DFD in Fig.~\ref{fig:example_3} contains a dangling arrow ``pol'': this is not an error in the figure, but rather an unfortunate side-effect of the way the original transformation rules were formulated (Fig.~\ref{fig:hotspots}). This and other shortcomings and inaccuracies are discussed at the end of Section~\ref{sub:transformation}.

\section{\uppercase{From B-DFD to PA-DFD}} \label{sec:algorithms}
\begin{figure}[tb]
\centering
\includegraphics[width=\linewidth]{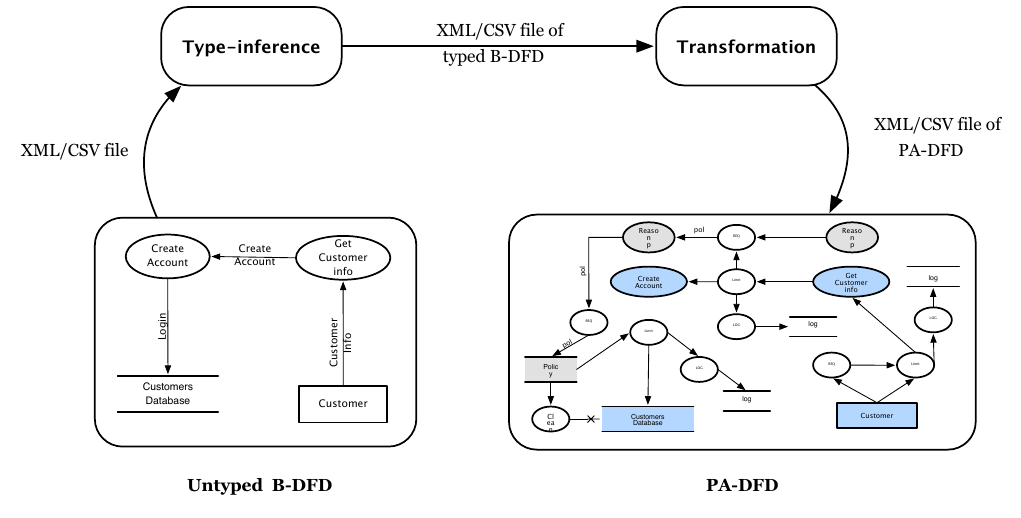}
\caption{A general architecture of the approach.}
\label{fig:general_architecture}
\end{figure}

\noindent We present here our algorithms for transforming \DFD to \PADFD.
The transformation process consists of two steps:
\emph{type-inference} followed by the actual \emph{transformation}.  Type-inference ensures that the input B-DFD is well-formed before it is transformed into a PA-DFD in the second step.  Fig.~\ref{fig:general_architecture} shows the general architecture of our approach.

\subsection{Type-inference}\label{sub:type-inference}

The B-DFDs we read from input files are not necessarily well-formed.
They may, for example, connect external entities directly, or contain a data deletion flow connecting two process entities rather than a process and a data store.
Such inconsistencies reveal errors made by designers.
Our tool detects and reports such issues.
For this purpose, we distinguish between two kinds of B-DFDs:
\emph{raw} B-DFDs correspond to diagrams read from input files and may contain inconsistencies;
\emph{well-formed} B-DFDs are free of inconsistencies and satisfy all
the necessary invariants required by our transformation algorithm.

We represent both kinds of B-DFDs as \emph{attributed multigraphs}
with activators as nodes and flows as edges.
Attributes allow us to specify properties of activators and flows,
such as their type or associated privacy information.

\begin{definition}
  \label{def0}
  An \emph{attributed multigraph} $G$ is a tuple $G = (\Activators, \Flows, \Attributes, \Values, s, t, \nAtt, \eAtt)$ where
  \Activators, \Flows, \Attributes and \Values are sets of nodes, edges, attributes and attribute values, respectively;
  $s,t\colon \Flows \rightarrow \Activators$ are the source and target maps;
  $\nAtt\colon \Activators \to (\Attributes \pto \Values)$ and $\eAtt\colon \Flows \to (\Attributes \pto \Values)$ are attribute maps that assign values for the different attributes to nodes and flows, respectively.
\end{definition}
\noindent Note that the attribute maps are partial, \ie nodes and edges may lack values for certain attributes.

In the following, we use the letters $n$, $m$ to denote nodes and $e$, $f$ to denote edges.  We write $\edgeT{e}{n}{m}$ to indicate that the edge $e$ has source $s(e) = n$ and target $t(e) = m$; we use ``.'' to select attributes, writing $n.a$ for $\nAtt(n)(a)$ and $f.a$ for $\eAtt(f)(a)$.
The set $\sourcesactivators(G) \subseteq \Activators$ of \emph{source nodes} in $G$ is defined as $\sourcesactivators(G) = \{ n \mid \exists e. s(e) = n \}$;
similarly, $\targetsactivators(G)$ denotes the set of \emph{target nodes} in $G$.

A (raw) B-DFD is simply an attributed multigraph with a fixed choice of attributes $\Attributes=\{ \type \}$.
The \type attribute describes the type of activators and flows.
Activators can be external entities (\externalentity), processes (\process) and data stores (\database);
flows are classified as either plain data flows (\pf) or data
  deletions (\df).
Fig.~\ref{fig:dfd} shows an example of a B-DFD with five activators (an external entity, a datastore and three processes) that are connected by plain flows.

\begin{definition}
  \label{def1}
  Define the set of \emph{data node types} as $\DataNodeTypes = \{\externalentity,\process,\database\}$
  and the set of \emph{raw flow types} as $\RawFlowTypes = \{ \pf, \df \}$.
  A \emph{(raw) B-DFD} is an attributed multigraph $G$ with activators as nodes and flows as edges, and where we fix \Attributes and \Values to be
  $\Attributes = \{\type\}$, $\Values = \DataNodeTypes \uplus \RawFlowTypes$.
  In addition, every activator and flow must have a type, \ie $n.\type \in \DataNodeTypes$ and $f.\type \in \RawFlowTypes$ must be defined for all $n$ and $f$.
\end{definition}
Since the \type attribute plays an important role in all DFDs, we introduce the following notation for typing activators (nodes) and flows (edges):
given an activator $n$, we write $\nodeT{n}{t}$ as a shorthand for $n.\type = t$; given a flow $\edgeT{f}{n}{m}$, we write $\edgeT[t]{f}{n}{m}$ to indicate that $f.\type = t$.

Well-formed B-DFDs differ from raw B-DFDs primarily in the choice of flow types.
Flows are typed based on their source and target activators (see the left-hand side of Fig.~\ref{fig:new_PADFD}).
Only some combinations of sources, targets and flow types are valid:
plain data flows (\pf) that carry data between processes are typed as \comp;
plain flows between external entities and processes are typed according to whether they collect data from an external entity (\into) or disclose data to an external entity (\out);
plain flows between processes to data stores either \store data (from process to store) or \readd data (from store to process); data deletions (\delete) always point from a process to a data store.
Flows that do not fall into one of these categories are \emph{ill-typed} and will be rejected by our type inference algorithm.
In addition to these flow typing constraints, we adopt some common rules from the DFD literature for well-formed B-DFDs:
diagrams may not contain loops (flows with identical source and target activators)
, activators cannot be isolated (disconnected from all other activators), and processes must have at least one incoming and outgoing flow~\citep[see e.g.][]{falkenberg1991,ibrahim2010formalization,dennis2018systems}.

\begin{definition}
  \label{def2}
  Define the set of \emph{data flow types} as $\DataFlowTypes = \{\into, \out, \comp, \store, \readd, \delete \}$.
  A \emph{well-formed B-DFD} is an attributed multigraph $G$, where
  $\Attributes=\{\type\}$ and $\Values= \DataNodeTypes \uplus \DataFlowTypes$.
  In addition,
  flows and activators are subject to the following conditions:
  \begin{itemize}
  \item $n.\type \in \DataNodeTypes$ and $f.\type \in \DataFlowTypes$;
  \item if $\edgeT[\into]{f}{n}{m}$ then $\nodeT{n}{\externalentity}$ and $\nodeT{m}{\process}$;
  \item if $\edgeT[\out]{f}{n}{m}$ then $\nodeT{n}{\process}$ and $\nodeT{m}{\externalentity}$;
  \item if $\edgeT[\comp]{f}{n}{m}$ then $\nodeT{n}{\process}$, $\nodeT{m}{\process}$ and $n \neq m$;
  \item if $\edgeT[\store]{f}{n}{m}$ then $\nodeT{n}{\process}$ and $\nodeT{m}{\database}$;
  \item if $\edgeT[\readd]{f}{n}{m}$ then $\nodeT{n}{\database}$ and $\nodeT{m}{\process}$;
  \item if $\edgeT[\delete]{f}{n}{m}$ then $\nodeT{n}{\process}$ and $\nodeT{m}{\database}$;
  \item if $\nodeT{n}{\process}$ then $\isSrc{n}$ and $\isTgt{n}$
  \item if $\nodeT{n}{\externalentity}$ or $\nodeT{n}{\database}$ then $\isSrc{n}$ or $\isTgt{n}$
  \end{itemize}
\end{definition}

The Type-inference algorithm (Algorithm~\ref{alg1}) detects and reports any ill-formed flows (lines~\ref{alg1:f1}--\ref{alg1:f2}) and activators (lines~\ref{alg1:a1}--\ref{alg1:a2}).
If type inference is successful, the resulting well-formed B-DFD can safely be transformed into a PA-DFD, as described in the next section.

\begin{algorithm}
  \Input{A raw B-DFD $G$}
  \Output{A well-formed version of $G$}
  \ForEach{$\edgeT{f}{m}{n} \in \Flows$ \label{alg1:f1}}{
    \If{$f.\type = \pf$}{
      \If{$\nodeT{m}{\externalentity}$ and $\nodeT{n}{\process}$}{
        $f.\type \gets \into$}
      \ElseIf{$\nodeT{m}{\process}$ and $\nodeT{n}{\externalentity}$}{
        $f.\type \gets \out$}
      \ElseIf{$\nodeT{m}{\process}$ and $\nodeT{n}{\process}$ and $m \neq n $}{
        $f.\type \gets \comp$}
      \ElseIf{$\nodeT{m}{\process}$ and $\nodeT{n}{\database}$}{
        $f.\type \gets \store$}
      \ElseIf{$\nodeT{m}{\database}$ and $\nodeT{n}{\process}$}{
        $f.\type \gets \readd$}
      \lElse{$f$ is ill-formed }
    }
    \ElseIf{$f.\type = \df$}{
      \If{$\nodeT{m}{\process}$ and $\nodeT{n}{\database}$}{
        $f.\type \gets $\delete}
      \lElse{$f$ is ill-formed}\label{alg1:f2}
    }
  }
  \ForEach{$n \in \Activators$ \label{alg1:a1}}{
    \lIf{ $\nodeT{n}{\process}$ and
      $(n \notin \setsources$ or $n \notin \settargets)$}{
      $n$ is ill-formed}
    \lElseIf{ $\nodeT{n} \externalentity$ and
      $(n \notin \setsources$ and $n \notin \settargets)$}{
      $n$ is ill-formed}
    \lElseIf{ $\nodeT{n}{\database}$ and
      $(n \notin \setsources$ and $n \notin \settargets)$}{
      $n$ is ill-formed}\label{alg1:a2}
  }
  \caption{Type-inference}\label{alg1}
\end{algorithm}

\subsection{Transformation}\label{sub:transformation}

\begin{figure}[tb]
\centering
\includegraphics[width=\linewidth]{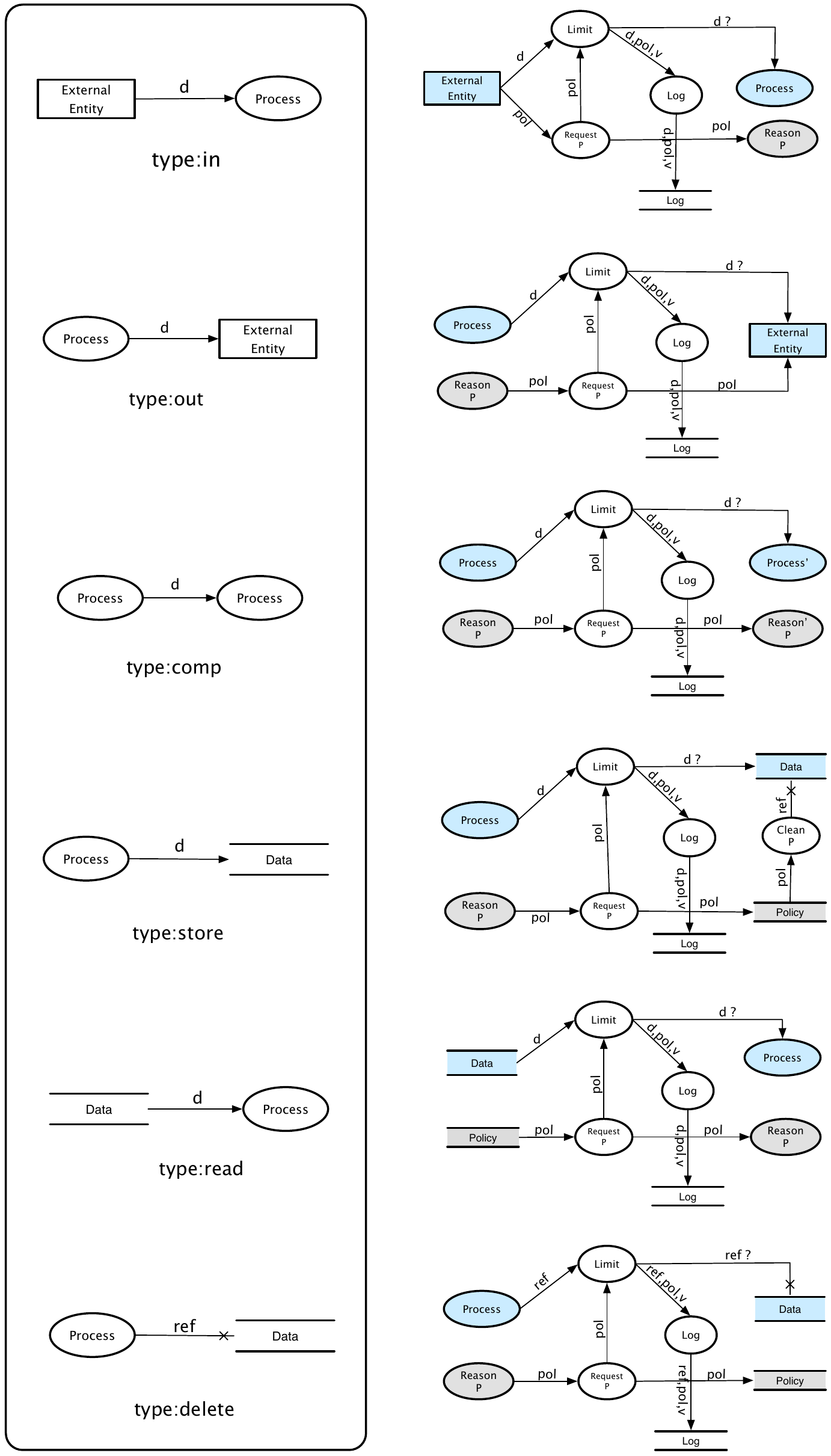}
\caption{Well-formed B-DFD and the updated corresponding PA-DFD elements.}
\label{fig:new_PADFD}
\end{figure}

Well-formed B-DFDs are guaranteed to be well-formed, but they do not yet contain any explicit privacy checks.
They are introduced by
Algorithm~\ref{alg2}, which transforms each flow in the well-formed B-DFD into a set of corresponding PA-DFD elements (see  Fig.~\ref{fig:new_PADFD}).
These PA-DFD elements represent the functionality necessary to enforce purpose limitation, retention time, accountability and policy management.

First we add \reason activators for each process in the well-formed B-DFD.
These activators are linked to each other by a special \partner attribute.
Each \reason activator is assigned to exactly one process via this attribute. Likewise, we add a new \policydb activator to each data store in the well-formed B-DFD and link them via their \partner attributes.
The second phase of the algorithm transforms each flow based on its type (\ie the hotspot that it belongs to). We use dedicated helper procedures to transform each flow type. For brevity, we only show the procedure \inaddnewelement, which transforms \into flows.  The others are similar. 
The auxiliary procedure \addnewcomelement is used to add elements that are common to all transformations.

Since PA-DFDs contain privacy checks, their activator and flow types are mainly different from DFDs (raw and well-formed). PA-DFDs have types of activators and flows that handle and carry policy called \emph{policy node types} and \emph{policy flow types},respectively. 
Besides the policy flow types, PA-DFDs also have types of flows that carry only data. For example, all the flows connecting limit activators to downstream activators are assigned as data flow type. Moreover, PA-DFDS have types of activators and flows that track and manage system events called \emph{admin node types} and \emph{admin flow types}. Note, the set of data node types is updated by including \limit since it is an activator that handles both data and policy. 
As with \DFD, we use attributed graphs to represent \PADFD formally.
\begin{definition}
  \label{def3}
  Define the set of \emph{data node types} as $\DataNodeTypes = \{ \externalentity,\process,\database,\limit \}$,
  the set of \emph{policy node types} as $\PolNodeTypes = \{\limit, \request, \reason, \policydb\}$, the set of \emph{admin node types} as $\AdminNodeTypes = \{\logg, \logdb, \clean\}$,
  the set of \emph{data flow types} as $ \DataFlowTypes= \{prolim, extlim, dblim, limpro,limext ,limdb,\\limdb\_del \}$,
  the set of \emph{policy flow types} as $ \PolFlowTypes =\{reqlim ,reqrea ,reqpdb ,reareq ,extreq ,reqext ,\\ pdbreq\} $ and
  the set of \emph{admin flow types} as $\AdminFlowTypes = \{ limlog, logging, pdbcle, cledb\_del \} $.
  A \emph{PA-DFD} is an attributed graph $G$, where
  $\Attributes = \{ \type, \partner \}$ and $\Values = \DataNodeTypes \uplus \PolNodeTypes \uplus \AdminNodeTypes \uplus \DataFlowTypesPA \uplus \PolFlowTypes \uplus \AdminFlowTypes \uplus \Activators$.
  In addition, flows and activators are subject to the following conditions:
  \begin{itemize}
  \item $n.\type \in \DataNodeTypes \uplus \PolNodeTypes \uplus \AdminNodeTypes$ ;
  \item $f.\type \in \DataFlowTypes \uplus \PolFlowTypes \uplus \AdminFlowTypes $;
  \item if $n.\partner$ is defined, then $n.\partner \in \Activators$.
  \end{itemize}
\end{definition}

In principle, the flows of PA-DFDs ought to be subject to similar typing conditions as those for well-formed B-DFDs.
Following the principle used for well-formed B-DFDs, we could type each flow based on the types of its source and the target activators.
For example, all the flows connecting request activators to limit activators could be assigned the type \emph{reqlim}.
This results in eighteen new flow types classified into three sets as present in Def.~\ref{def3}. 

\begin{algorithm}
  \Input{A well-formed B-DFD $G$}
  \Output{A PA-DFD }
  \ForEach{ $n \in \Activators$}{
    \If{$\nodeT{n}{\process}$}{
      add a new node $\nodeT{m}{\reason}$ to $G$;\\
      $m.\partner \gets n$; $n.\partner \gets m$}
    \If{$\nodeT{n}{\database}$}{
      add a new node $\nodeT{m}{\policydb}$ to $G$;\\
      $m.\partner \gets n$; $n.\partner \gets m$}
  }
  \ForEach{ $f \in \Flows$}{
    \lIf{$\nodeT{f}{\into}$}{
      \inaddnewelement($n,f,G$)
    }
    \lIf{$\nodeT{f}{\out}$}{
      \outaddnewelement($n,f,G$)
    }
    \If{$\nodeT{f}{\comp}$}{
      \compaddnewelement($n,f,G$)
    }
    \lIf{$\nodeT{f}{\store}$}{
      \storeaddnewelement($n,f,G$)
    }
    \lIf{$\nodeT{f}{\readd}$}{
      \readaddnewelement($n,f,G$)
    }
    \If{$\nodeT{f}{\delete}$}{
      \deleteaddnewelement($n,f,G$)
    }
  }
  \caption{Transformation}\label{alg2}
\end{algorithm}

\begin{procedure}
  add a new activator $\nodeT{n_0}{\limit}$ to $G$;\\
  add a new activator $\nodeT{n_1}{\request}$ to $G$;\\
  $n_0.\partner \gets n_1$; $n_1.\partner \gets n_0$\\ 
  add a new activator $\nodeT{n_2}{\logg}$ to $G$;\\
  add a new activator $\nodeT{n_3}{\logdb}$ to $G$;\\
  add a new flow $ \edgeT[reqlim]{f_{0}}{n_1}{n_0}$ to $G$;\label{common_reqlim}
  \\
  add a new flow $ \edgeT[limlog]{f_{1}}{n_0}{n_2}$ to $G$;
  \\
  add a new flow $ \edgeT[logging]{f_{2}}{n_2}{n_3}$ to $G$;
  \\
  \caption{addCommonElems($f$,$G$)}\label{proc:common}
\end{procedure}

\begin{procedure}
  $n_0,n_1,n_2,n_3 \gets$ \addnewcomelement($f, G$)
  \\
  add a new flow $\edgeT[extlim]{f_3}{s(f)}{n_0}$ to $G$;
  \\
  add a new flow $\edgeT[extreq]{f_4}{s(f)}{n_1}$ to $G$;
  \\
  $f_3.\partner \gets f_4$; $f_4.\partner \gets f_3$;\label{procin:p1}
  \\
  add a new flow $\edgeT[reqrea]{f_5}{n_{1}}{n.\partner}$ to $G$;
  \\
  $f.\type \gets limpro$; $s(f)\gets n_{0}$;
  \\
  $f.\partner \gets f_5$; $f_5.\partner \gets f$;\label{procin:p2}
  
  \caption{addInElems($n$,$f$,$G$)}\label{proc:in}
\end{procedure}

\subsubsection{Comparison of transformation rules}
The transformation rules presented in Fig.~\ref{fig:hotspots} have a few subtle but important shortcomings that are addressed in our Transformation algorithm.

First---and most importantly---the rules do not explain how activators with multiple input and output flows are to be transformed.  Indeed, the activators in the left-hand sides of the rules have at most one incoming or outgoing flow.  This begs the question which of the newly introduced privacy activators (and flows) are to be added only once per rule application, and which need to be instantiated for every separate incoming (or outgoing) flow.
We solve this problem by splitting the transformation into two distinct steps. In a first step, we create \emph{partners} for process and data store activators: each process receives a new \emph{reason} node, and each data store a \policydb, as its partner.
In the second step, each flow is equipped with \emph{limiting}, \emph{requesting} and \emph{logging} activators and the necessary flows to connect the newly created nodes to the original activators. This two-step approach gives a clean separation of the per-activator and per-flow aspects of each rule.

Second, the limiting and logging activators in the original rules are set up in a semantically dubious way. Every \limit activator is followed by a \logg activator that receives both a policy and a data value. The \logg activator logs both values and forwards the data value to downstream activators (such as processes, external entities or data stores).  But the purpose of the \limit activator is to inhibit unintended flows that would cause privacy violations, so it must only pass on policy-compliant data values. This means that policy violation events never reach the \logg activator, and are therefore not logged. This seems highly problematic.  An alternative interpretation is that \emph{all} data and policy values (irrespective of potential violations) are passed from the \limit to the \logg activator to avoid this, and that the \logg activator performs the actual filtering. But why have separate \limit and \logg activators in the first place then? We resolve this ambiguity by connecting \limit activators directly to the downstream activators, and then separately to the \logg activator in charge of registering possible policy violation. To this end, the flow connecting the \limit and \logg activators carries a special flag $v$ indicating whether a violation took place (see the right-hand side of Fig.~\ref{fig:new_PADFD}).

\begin{figure}
\centering
\includegraphics[width=0.99\linewidth]{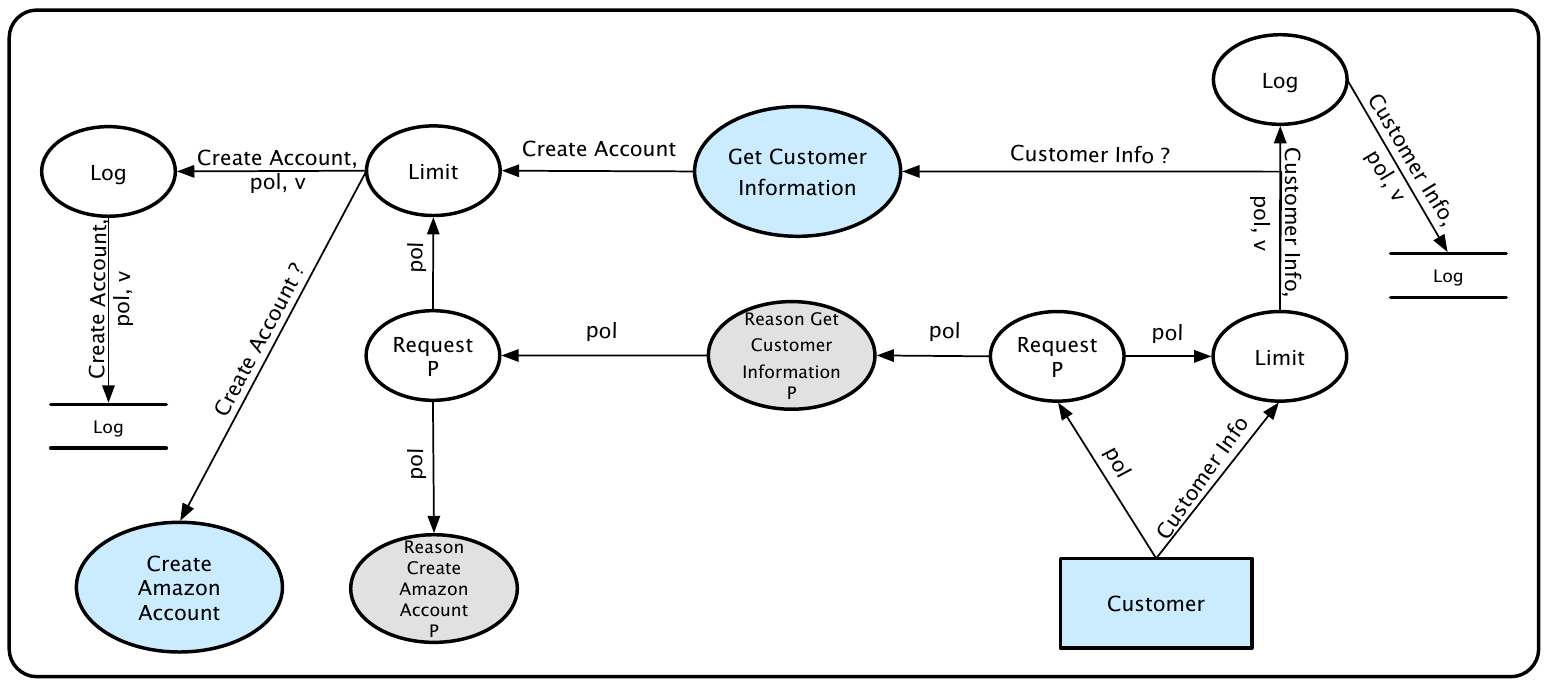}
\caption{Example of a PA-DFD generated by the updated rules.}
\label{fig:example_4}
\end{figure}

Finally, the original ``Usage'' rule contains a subtle error, which is again related to the way it connects the newly introduced \limit and \logg activators (see the right-hand side of Fig.~\ref{fig:hotspots}). After the application of this rule, the process activator receives an additional policy value $pol$ in addition to the data value $d$ it received originally. It passes this value to the \logg activator immediately succeeding it, along with the updated data value $d'$. One has to assume that the process does so without changing the value of $pol$. Otherwise, how would the \logg activator know the original policy (which may have been used by the preceding \limit activator to detect a privacy violation)? But this means that the data value $d'$ and the policy value $pol$ are out of sync. Fig.~\ref{fig:example_3} shows an example of PA-DFD that is transformed according to the original rules. In this example, there are two hotspots ``Collection'' and ``Usage''. The ``Get Customer Information'' process has input and output flows labelled ``Customer info'' and ``Create Account'', respectively, which matching the pattern of a ``Usage'' hotspot. As a result, it is transformed according to the ``Usage'' rule as shown in Fig.~\ref{fig:example_3} where the aforementioned subtle errors appear more clearly. For instance, the ``Get Customer Information'' process receives the ``Customer Info'' and the corresponding policy ``pol'', then passes it to the
``Log'' activator with the processed data ``Create Account''. This means there is a mismatch between the logged data ``Create Account'' and the policy value ``pol''. Indeed, it is unclear what exactly is supposed to be logged by the \logg activator. The outcome of the \limit activator or that of the process?  Our algorithm removes this ambiguity by separating limiting and logging, which are added on a per-flow basis, from processing. As a result, there are two separate \limit activators (each with its own \logg activator) protecting the input and output flows of the process, and the process never receives a policy value.
Fig.~\ref{fig:example_4} shows the PA-DFD produced by our Transformation algorithm for the same B-DFD.
Note that the ``Create Account'' flow, after having been transformed according to the \comp rule, is protected by its own \limit and \logg activators, and there are no longer any dangling flows.

PA-DFD preserves the functionality of all flows in the original DFD if they do not lead to any potential privacy violations. By functionality, we mean the activators of the original DFD and their connective structure. Since PA-DFD contains privacy checks to inhibit unintended flows that would cause privacy violations, some flows do not pass carried data values. For example, data flows that connect \limit nodes directly to the downstream activators only pass on policy-compliant data values.

\subsection{Our Tool}\label{sub:implementation}
We have modified the hotspots-based translation given by \cite{ASS16pac,ASS18pcm} in order to address its ambiguities and inaccuracies. Our tool for transforming \DFD into \PADFD implements algorithms \ref{alg1} and \ref{alg2}, and uses a third-party application for drawing the diagrams. Such drawing software should support the drawing of DFDs, be user-friendly, be easy-to-use, be cross-platform and be open source. \texttt{draw.io} was the tool of our choice \citep{drawio}. Although there is no default library for \DFD included in \texttt{draw.io}, users can easily install Henriksen's custom library which has all the elements that are needed for B-DFD. The library is hosted in a public Github repository \citep{Michaellibraries}.
Since it is easy to import and export diagrams from/to XML format in \texttt{draw.io}, our implementation processes B-DFD diagrams represented in an XML format and generates PA-DFD diagrams in the same format.

Our tool is implemented in Python and has been tested on a MacBook Pro.\footnote{The source code of our tool is available at\\
  \gitrepo{}.}

\section{\uppercase{Case Studies}}\label{sec:casestudy}

\noindent
To show the feasibility of our approach and validate our algorithms, we have applied our tool to models of two realistic applications:
an automated payment system and an online retail system.
We illustrate the correctness of our algorithm by running informal simulations of the two models.

\subsection{Automated Payment System}

The DFD for the secure payment system considered here is due to \cite{chong2020integrating};
it has been reviewed by domain experts and
models a system for making automatic payments to subcontractors in a construction project.

\begin{figure}[tb]
\centering
\includegraphics[width=0.80\linewidth]{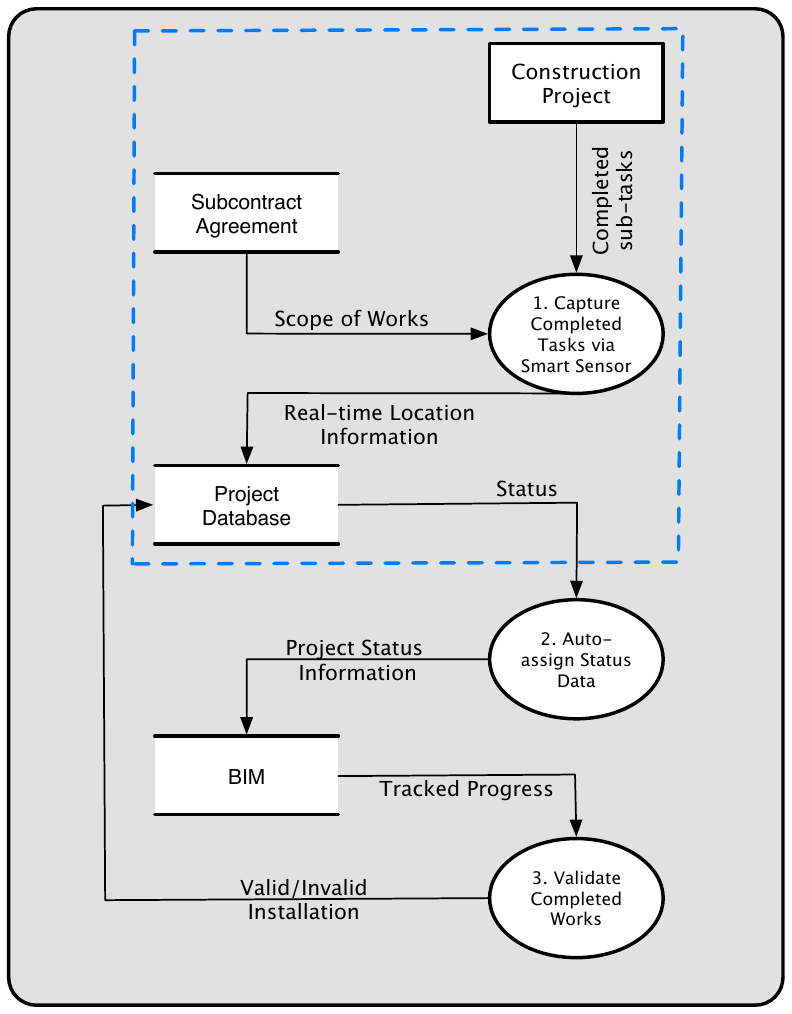}
\caption{Part of Automated Payment System DFD.}
\label{fig:BDFD_CS1}
\end{figure}

As shown in Fig.~\ref{fig:BDFD_CS1}, the automated payment system consists of the processes
numbered 1--3.
Process~1 recognises finished sub-tasks via a smart sensor located at the construction site.
The process receives two inputs:
``Completed sub-tasks'' and ``Scope of Works''; the latter describes the subcontractors' and suppliers' contractual duties and is required to determine the type of information captured by the smart sensor. The process transfers information about material location and performance gathered by the smart sensor to the project data base.
Process~2 automatically assigns up-to-date information from the project data store to the ``Building Information Modelling'' (BIM) data store. BIM is a method for designing and managing information about a construction project during its entire life cycle. The ``Status'' flow represents updates generated when the project data store receives and stores new data from the smart sensor.
Process~3 validates completed sub-tasks following quality requirements; it keeps the project database up-to-date according to the ``Tracked Progress'' information from the BIM DB by storing and marking each sub-task as a ``Valid/Invalid Installation'' in the project database.
For the complete DFD and further details see \cite{chong2020integrating}.

\subsubsection{Informal simulation of the PA-DFD}

To evaluate our approach and increase confidence in its correctness, we perform an informal simulation of the payment system, illustrating that the PA-DFD generated by our proof-of-concept tool enforces the desired GDPR properties (purpose limitation and accountability).
We start by augmenting the original DFD with \emph{static (or design-time)} policy information.
Table~\ref{tab:Static_info} shows an extract: each data flow is assigned a unique identifier (\emph{F\_id}), its \emph{Label} (from the DFD),
a \emph{Purpose}
(to be checked against the data subject consent),
a \emph{PD} flag indicating whether the piece of data contains personal information, and a \emph{Data\_type}
(\eg ``image'' and ``email'').

\begin{figure}[tb]
\centering
\includegraphics[width=0.80\linewidth]{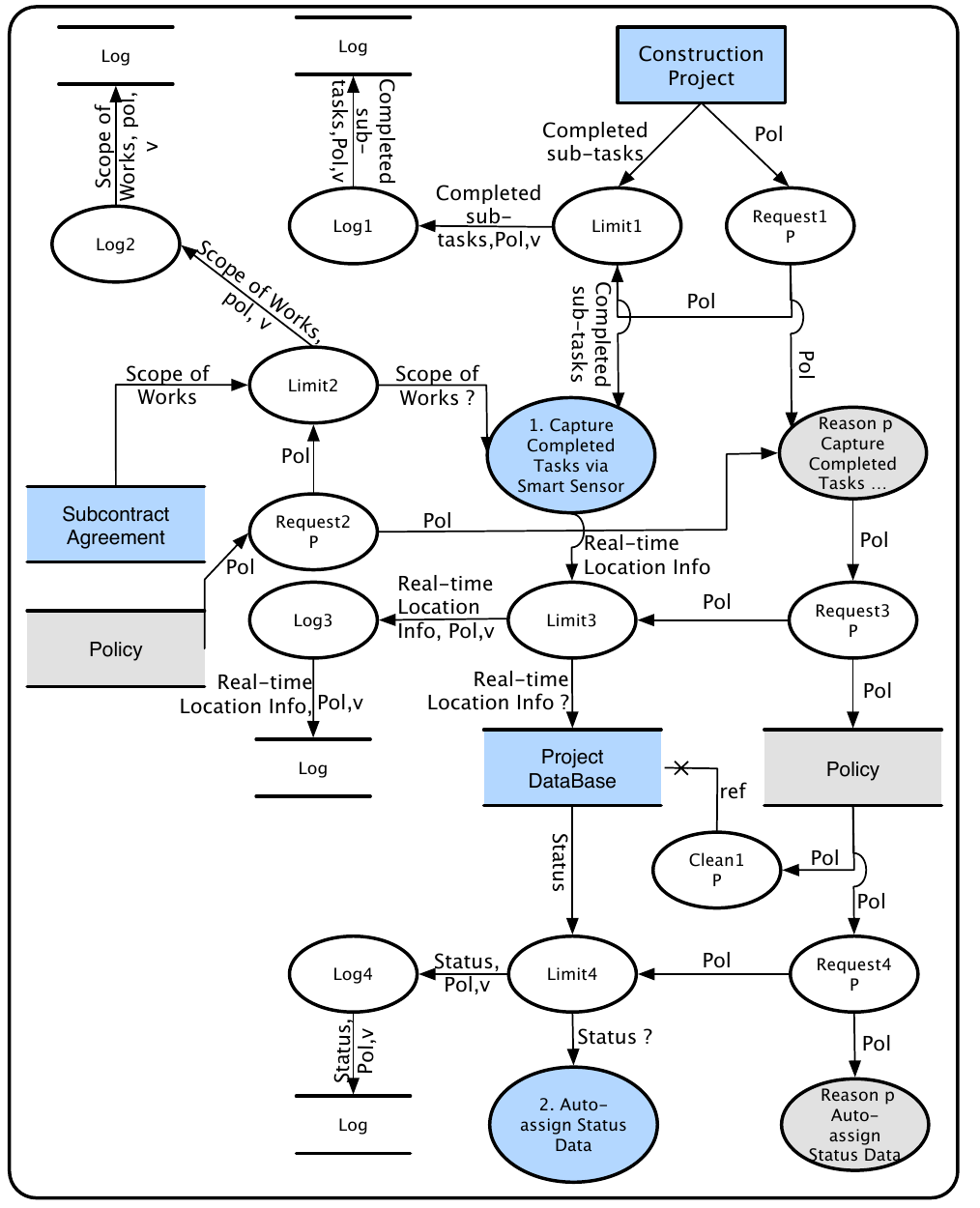}
\caption{Part of Automated Payment System PA-DFD.}
\label{fig:PADFD_CS1}
\end{figure}

Next, we transform the B-DFD thus obtained into a PA-DFD, parts of which are shown in Fig.~\ref{fig:PADFD_CS1}.
To run the informal simulation, we assume a set of \emph{dynamic} information provided by users during runtime. An excerpt is shown in Table~\ref{tab:Dynamic_info}.  Each row consists of a unique data identifier (\emph{D\_id}) with five attributes: \emph{F\_id} indicates the flow that carries the user data; \emph{DSub} is the data subject; \emph{Pol/Consent} is a set of consented purposes; \emph{Expiry} defines the expiration time for the data; \emph{Content} is the actual data.
The last two columns of Table~\ref{tab:Dynamic_info} indicate whether the given data values are forwarded to downstream activators in the B-DFD and PA-DFD, respectively, during the simulation.
They illustrate that the PA-DFD prevents some data values from being processed, while the B-DFD forwards all data values to downstream activators for processing, storing, or reading regardless of the data subject privacy preferences since there are no privacy checks.

Consider, for example, the ``Completed sub-tasks'' flow between ``Construction Project'' and Process~1 in the original DFD.  This flow carries sensitive information collected by the smart sensor, which needs to be checked and limited according to the subcontractor's privacy policy.
This is achieved via corresponding \limit and \request activators in the PA-DFD.  These enforce the principles of purpose limitation and the lawful ground of consent.  To illustrate this we consider two scenarios, represented by the data values $d_1$ and $d_5$ in the first and last rows of Table~\ref{tab:Dynamic_info}, respectively. In the first scenario, we assumed the data subject ``SubcontractorX'' permitted the smart sensor to collect information
until the end of 2020. Consequently, the information $d_1$ is forwarded from the \limit node to Process~1 and logged (in accordance with the accountability principle) in the \logg store along with the corresponding policy and a flag indicating that no violation occurred.  In the second scenario, the \limit node prevents the data value $d_5$ from being forwarded to Process~1 since the intended purpose of the flow (``Capturing completed sub-tasks'') is not compatible with the purpose (``Taking pictures for advertisements'') to which the data subject (``SubcontractorY'') consented.
Furthermore, this event is logged in the \logg store and identified as a privacy violation.

Contrast the above scenarios with the original DFD in which the subcontractors' data is unconditionally forwarded to processes and stored without regard for any GDPR principle; the data can be collected and used without any limitation for any purpose.
For example, the data of ``SubcontractorY'' ($d_5$) is collected and processed, even though they might not have been consented to this particular use, potentially violating privacy.

\begin{table*}
\centering
\caption{Design Time Information for B-DFD Automated Payment System.}
\label{tab:Static_info}
\resizebox{\textwidth}{!}{
\begin{tabular}{|c|c|c|c|c|}
  \hline
  F\_id & Label & Purpose & PD & Data\_type \\
  \hline
  $f_1$ & Completed sub-tasks & Capturing completed sub-tasks & True & video, images and string \\
  $f_2$ & Scope of Works & Knowing subcontractors contractual duties & True & string \\
  $f_3$ & Real-time Location Information & Project monitoring & True & video, images and string \\
  $f_4$ & Status & Sending up to date project information to IBM & True & video, images and string \\
  \hline
\end{tabular}}
\end{table*}

\begin{table*}
\centering
\caption{Run-time Information for B-DFD/PA-DFD Automated Payment System.}
\label{tab:Dynamic_info}
\resizebox{\textwidth}{!}{
\begin{tabular}{|c|c|c|c|c|c|c|c|}
  \hline
  D\_id & F\_id & Dsub & Pol/Consent & Expiry & Content & Fwd.~in~B-DFD &Fwd.~in~PA-DFD \\
  \hline
  $d_1$ & $f_1$ & SubcontractorX & Capturing completed sub-tasks & end of 2020 & "streaming videos" and "image\_1.jpg" & Yes & Yes \\
  $d_2$ & $f_2$ & SubcontractorX & Identifying assigned tasks & end of contract & "facade panel installation" & Yes &  Yes \\
  $d_3$ & $f_3$ & SubcontractorX & Recording the work status & end of contract & "streaming videos" and "image\_2.jpg" & Yes &  Yes \\
  $d_4$ & $f_4$ & ProjectX & Assigning project info to BIM & end of 2021 &"Project info:name, desc, status, subcontract,etc" &Yes& Yes \\
  $d_5$ & $f_1$ & SubcontractorY &  Taking pictures for advertisements  & end of 2020 & "streaming videos" and "image\_3.jpg" & Yes & No  \\
  \hline
\end{tabular}}
\end{table*}

\subsection{Online Retail System}

\noindent
As a second case study, we have applied our tool to a high-level B-DFD model of online retail sector system.\footnote{\url{https://creately.com/diagram/iusq4h6z1/Amazon\%20level-0\%20DFD}}.  This case study represents a publicly available, real-world use case of DFDs in a context where privacy is highly relevant; it is known that privacy issues abound in online retailing \citep{kuriachan2014online}.

\begin{figure}[tb]
\centering
\includegraphics[width=0.80\linewidth]{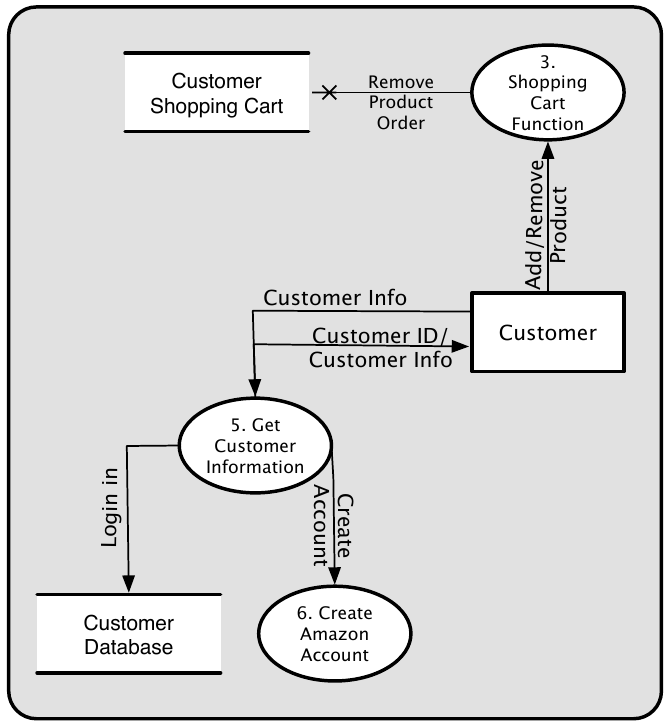}
\caption{Part of B-DFD of Online Retail Order System.}
\label{fig:DFDAmazon}
\end{figure}

Fig.~\ref{fig:DFDAmazon} shows a part of the B-DFD of the online retail model.
It contains activators for processing orders (\eg ``Shopping Cart Function'') and for managing customer account information (\eg ``Get Customer Information'' and ``Create Account'').

\begin{figure}[tb]
\centering
\includegraphics[width=0.98\linewidth]{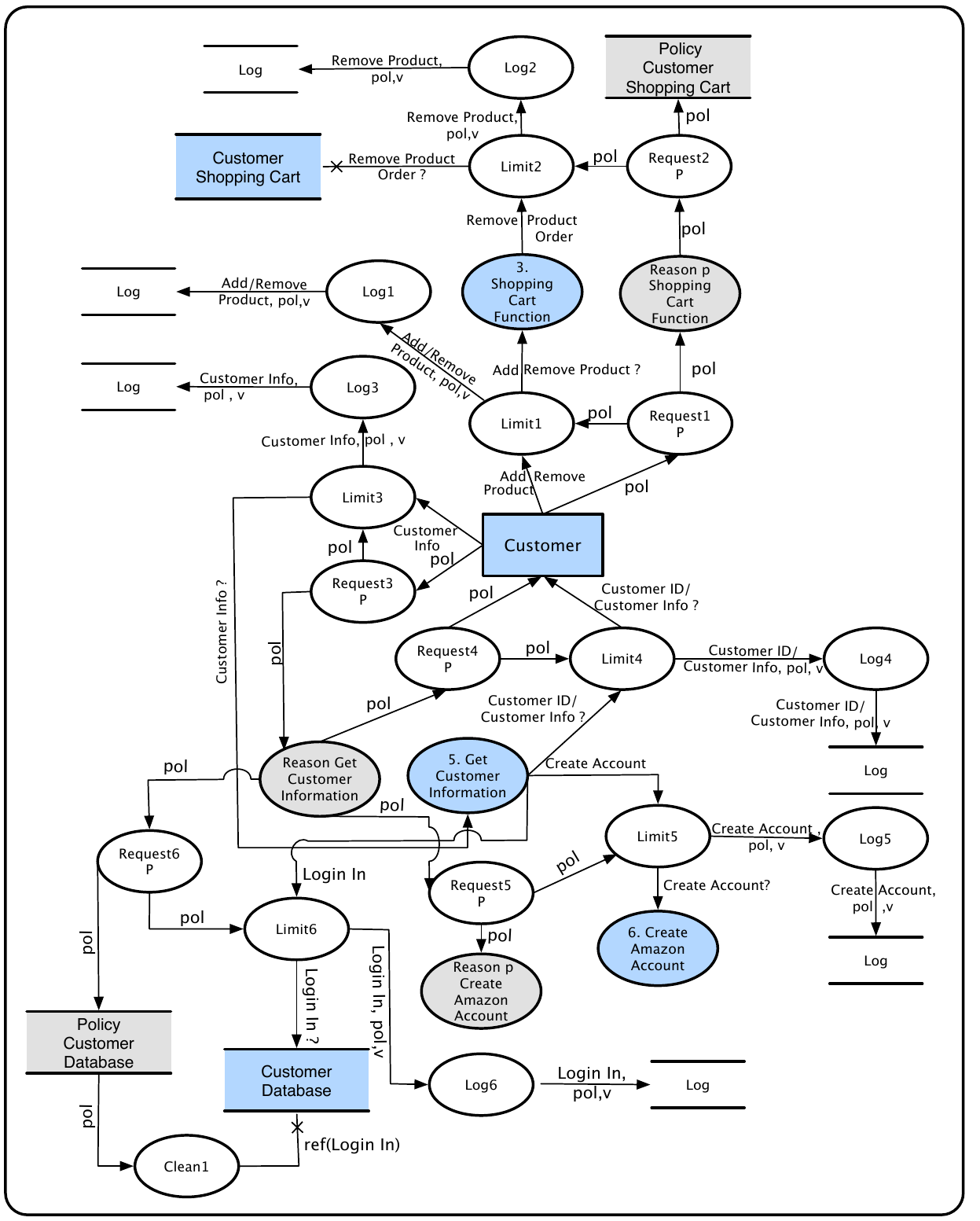}
\caption{Part of PA-DFD of Online Retail Order System.}
\label{fig:PA-DFDAmazon}
\end{figure}

Fig.~\ref{fig:PA-DFDAmazon} shows a part of the produced PA-DFD of an Online retail Order System. The tool typed all the flows and produced the expected transformation. The added privacy activators indicate to engineers where privacy checks have to be implemented.
For example,
the flow between ``Customer'' and ``Get Customer Information'' that carries personal data needs to be checked and limited according to the customer's
privacy policy.
This is achieved, in the PA-DFD, via the corresponding \limit and \request activators.
By contrast, the original online retail B-DFD allows a customer's data to be processed and stored
in ways that infringe on her privacy
because there is no specification of how the customer expects her data to be used (consent), nor any privacy checks enforcing that specification, \ie that the data is actually used for the purpose for which it was collected (purpose limitation and data minimization).

\section{\uppercase{Related work}}
\label{sec:relatedw}

\noindent Though PbD has been advocated since the 1990s, its use in realistic systems is relatively recent.
Some examples include electronic traffic pricing (ETP), smart metering, and location-based services \citep[see][and references therein]{LeMetayer13pbd}.
Our approach to PdB builds on the work by \cite{ASS16pac,ASS18pcm}, which we discuss in detail in Section~\ref{subsec:privacy_DFDs}.

According to \cite{AM14pbd}, previous work on PbD has ``focused
on technologies rather than methodologies and on components rather than architectures''.  They
propose a more formal approach to PbD 
based on a {\em privacy epistemic logic} for specifying and proving privacy properties.

Another line of work in PbD is to consider privacy not from the architectural point of view but at a higher level of abstraction. For instance, \citet{Hoepman14pds} proposes privacy {\em design strategies} to be taken into account already during the requirement elicitation phase, long before designing the software architecture.
Along the same lines, \cite{CHH16cap} consider an additional level of abstraction between privacy design strategies and privacy patterns by considering {\em tactics}, and
\cite{NCK+14} present a methodology for engineering privacy based on existing state-of-the-art approaches like privacy impact assessment and risk management, among others.
By taking privacy into account even before the concrete design phase, these approaches allow software engineers to identify potential privacy issues early on.
They are complementary to our approach, which is at the model level and closer to the implementation.

\cite{basin2018purpose} have recently proposed a methodology to audit GDPR compliance by using {\em business process models}.
They identify ``purpose'' with ``process'' and show how to automatically generate privacy policies and detect violations of data minimisation at the modelling level.
The paper highlights the difficulty of representing the notion of \emph{purpose} at the programming language level, and provides convincing arguments on why GDPR compliance cannot be entirely automated.

\cite{SRK+18} present a definition of rules for achieving {\em Confidentiality-by-Construction}.  In their approach, functional
specifications
are replaced by confidentiality specifications listing which variables contain secrets.
Though the approach seems interesting, it has (to the best of our knowledge) not been fully implemented.

\citet{TSB19fif} propose an approach to analyse information flow (security) policies at the modelling level.  They focus on data confidentiality and integrity, and introduce a graphical notation based on DFDs to algorithmically detect design flaws ``in the form of violations of the intended security properties''. The approach has been implemented as a plugin for Eclipse and evaluated on real-world case studies.  While our work and theirs are both based on DFDs, the objectives are different: we focus on the implementation of automatic model transformation for specific privacy checks (retention time and purpose limitation), while \citeauthor{TSB19fif}~focus on the detection of design flaws associated with the security properties.

Our paper distinguishes itself
in that none of the above
has taken the approach to automatically add privacy checks to design models.

Finally, for further discussion on the challenges and open problems of PbD and privacy-by-construction, see \cite{TBC15pbd,schneider18pbc} and the comprehensive ENISA report by \cite{enisa15pdp}.

\section{\uppercase{Conclusions}}
\label{sec:conc}

\noindent We have provided algorithms to automatically translate DFD models into privacy-aware DFDs (PA-DFDs) as well as a proof-of-concept implementation integrated into a graphical tool for drawing DFDs. This paper is the practical realisation of previous work that only presented the idea of enhancing DFDs with privacy checks and a very high level transformation between both models. Obtaining the algorithms (from the existing conceptual transformation) was not a straightforward task as some aspects of the transformation were more subtle than expected and some of the intuitions
underlying
the high-level graphical transformation turned out to be flawed.
We have addressed these conceptual flaws in our algorithms and evaluated them through two case studies: an automated payment system
and an
online retail system.

One limitation of our approach concerns the readability of the PA-DFD: the diagrams resulting form our transformation can be large, making it difficult to visualise them. That said,
the intended use of this tool is as an intermediate step in the design and development process.
Ideally, a software architect should have to inspect (and possibly modify) only small, relevant subsets of the PA-DFD.
Our next step is to implement an algorithm to automatically synthesise a template from the PA-DFD. Such a template should contain code skeletons (in Java, Python, etc.) for the basic components of the original (functional) DFD as well as the privacy checks from the PA-DFD.
We also intend to provide the programmer with pre-defined libraries to be used as building blocks for implementing such privacy checks.

\section*{\uppercase{Acknowledgements}}
This research has been partially supported by the Cultural Office of
the Saudi Embassy in Berlin, Germany and by the Swedish Research
Council ({\it Vetenskapsr\aa det}) under Grant~2015-04154 ``PolUser''.

\bibliographystyle{apalike}

\end{document}